\documentclass[prl,twocolumn,superscriptaddress,floatfix]{revtex4-2}
\usepackage{graphicx}
\usepackage{latexsym}
\usepackage{amsmath}
\usepackage{amsfonts}
\usepackage{amssymb}
\usepackage{bm}
\usepackage{txfonts}
\usepackage[markup=default]{changes}
\newcommand{\fig}[2]{\includegraphics[width=#1]{#2}}
\def\bq{{\textbf{q}}}
\def\bk{{\textbf{k}}}
\def\br{{\textbf{r}}}
\usepackage[colorlinks = true, linkcolor = blue, urlcolor  = blue, citecolor = blue, anchorcolor = blue]{hyperref}
\begin{document}
\title{\texorpdfstring{Nodeless ${\bm d}$-wave superconductivity driven by charge fluctuations in monolayer FeSe on SrTiO$_{\bm 3}$}{Nodeless d-wave superconductivity driven by charge fluctuations in monolayer FeSe on SrTiO3}}

\author{Jin-Wei Dong}
\affiliation{Anhui Province Key Laboratory of Condensed Matter Physics at Extreme Conditions, High Magnetic Field Laboratory, Chinese Academy of Sciences, Hefei 230031, China}

\author{Sen Zhou}
\thanks{Contact Author: zhousen@itp.ac.cn}
\affiliation{Institute of Theoretical Physics, Chinese Academy of Sciences, Beijing 100190, China}
\affiliation{School of Physical Sciences, University of Chinese Academy of Sciences, Beijing 100049, China}

\author{Jian Wang}
\affiliation{International Center for Quantum Materials, School of Physics, Peking University, Beijing 100871, China}

\author{Gang Su}
\affiliation{Institute of Theoretical Physics, Chinese Academy of Sciences, Beijing 100190, China}
\affiliation{School of Physical Sciences, University of Chinese Academy of Sciences, Beijing 100049, China}

\author{Ziqiang Wang}
\thanks{Contact Author: wangzi@bc.edu}
\affiliation{Department of Physics, Boston College, Chestnut Hill, MA 02467, USA}

\date{\today}

\begin{abstract}
Monolayer FeSe and Fe(Te,Se) grown on SrTiO$_3$ exhibit some of the highest superconducting transition temperatures among iron-based superconductors, yet their pairing mechanism and symmetry remain elusive.
Motivated by the recent surprising discovery of nodeless $d$-wave superconductivity using scanning Josephson tunneling microscopy (SJTM) [T. Wei \textit{et al.}, arXiv:2610.38421 (2026)], we investigate superconductivity in monolayer FeSe/SrTiO$3$ (1L-FeSe/STO) within an extended Hubbard model incorporating both onsite and intersite Coulomb interactions among Fe $3d$ and Se $4p$ orbitals.
We show that Fe-Se intersite Coulomb repulsion strongly enhances small-momentum $p$-$d$ charge-transfer fluctuations, which mediate attractive intraorbital spin-singlet pairing interactions.
These electronic fluctuations may be further enhanced by forward-scattering optical phonons of the SrTiO$3$ substrate.
The resulting intrapocket attraction, together with residual repulsive interpocket interactions, stabilizes a mirror-odd, nodeless $B_{1g}$ ($d$-wave) superconducting state with opposite gap signs on the two electron pockets.
Both Fe $t_{2g}$ and Se $4p$ orbitals contribute substantially to this multiorbital superconductivity.
The calculated superconducting gap and density of states exhibit a two-peak structure, while the Josephson tunneling conductance displays a half-lattice-period modulation, in excellent agreement with experimental observations.
Our results establish electronic charge-transfer fluctuations as a viable pairing mechanism for unconventional nodeless $d$-wave superconductivity in monolayer FeSe/STO.
\end{abstract}
\maketitle

\textit{Introduction.} - 
The discovery of monolayer FeSe grown on SrTiO$_3$ (1L-FeSe/STO) is one of the most striking developments in the field of iron-based superconductors (FeSC) \cite{XueQK-CPL12, ZhouXJ-NC12, ZhouXJ-NM13, FengDL-NM13, MaXC-CPL14, FengDL-PRL14, JiaJF-NM15, WangYY-SB15, ZhouXJ-NC21}. Despite exhibiting the highest superconducting (SC) transition temperatures $T_c$ among all FeSCs, the pairing mechanism and SC order parameter symmetry remain unresolved \cite{ZhouXJ-JPCM15, Hoffman-ARCMP17, LeeDH-CPB15, LiZX-SB16}. In bulk FeSCs, superconductivity is widely believed to have $s$-wave ($A_{1g}$) pairing symmetry, with an $s_\pm$ gap structure characterized by opposite signs on the electron and hole Fermi pockets, as naturally favored by repulsive interactions associated with antiferromagnetic spin fluctuations \cite{Mazin-PRL08,Kuroki-PRL08}.
In contrast, 1L-FeSe/STO appears heavily electron-doped and possesses only electron pockets near the Brillouin-zone corners, while the hole bands at the zone center lie below the Fermi level \cite{ZhouXJ-NC12,ShenZX-PRL16,FengDL-NC14,BaoX-NC17,Zhao-SA18}.
The absence of hole pockets challenges the electron-hole sign-changing $s$-wave pairing scenario and raises fundamental questions about the pairing interaction and symmetry responsible for its high-temperature superconductivity.

Experimentally, angle-resolved photoemission spectroscopy (ARPES) and scanning tunneling microscopy (STM) have established a fully opened, moderately anisotropic SC gap on the electron pockets of 1L-FeSe/STO \cite{ShenZX-PRL16, FengDL-NP15}.
However, these measurements primarily probe the gap magnitude and cannot determine its relative phase, leaving the pairing symmetry under debate \cite{FengDL-NP15, XueQK-NC23,Ge-NL19,WangJ-PRL19, LiL-CP20, WangQH-PRL18}.
Various theoretical proposals have been advanced, including sign-preserving $s_{++}$ and incipient $s_\pm$ gap structures with $A_{1g}$ ($s$-wave) symmetry, as well as sign-changing pairing between the two electron pockets \cite{LeeDH-PRB13, Kang-PRL16, WangQH-PRB16, Chubukov-PRL12, Kontani-PRL10, Kontani-PRL12, Kontani-PRB17, Hirschfeld-PRB15, Hirschfeld-PRL16}.
Particularly intriguing is the proposal of nodeless $d$-wave superconductivity by Agterberg \textit{et al.} \cite{Agterberg-PRL17}. 
In this scenario, the SC order parameter changes sign between the electron pockets centered at $(\pi,0)$ and $(0,\pi)$ in the unfolded Brillouin zone, while remaining fully gapped because the pockets avoid the symmetry-enforced nodal directions. Agterberg \textit{et al.} further showed that this nodeless $d$-wave state can remain robust against electron-pocket hybridization when the relevant spin-orbit coupling (SOC) is sufficiently weak compared with the SC pairing scale. Thus, a fully opened gap does not necessarily imply $s$-wave pairing symmetry in monolayer FeSe.

Remarkably, recent scanning Josephson tunneling microscopy (SJTM) experiments by Wei \textit{et al.} \cite{SJTM} have revealed spatial Josephson interference patterns that support nodeless $d$-wave pairing in monolayer Fe(Te,Se)/STO and FeSe/STO, while reproducing the characteristic interference pattern of nodal $d$-wave pairing in the cuprates. 
Using an $s$-wave SC tip, the experiments revealed a striking half-lattice-period modulation of the Josephson tunneling signal, with maxima near the midpoints of nearest-neighbor Fe-Fe bonds and strong suppression along next-nearest-neighbor Fe-Fe directions. 
Together with the fully opened SC gap observed in single-particle tunneling spectra, this spatial Josephson interference pattern is consistent with a sign-reversing, nodeless $B_{1g}$ ($d$-wave) order parameter. 
These surprising observations provide strong motivation to revisit the microscopic origin of superconductivity in monolayer FeSe and, in particular, to identify pairing interactions capable of stabilizing a fully gapped $d$-wave state.

Most theoretical studies of superconductivity in 1L-FeSe/STO have focused on spin or orbital fluctuations, often supplemented by forward-scattering optical phonons of the SrTiO$_3$ substrate \cite{LeeDH-PRB13, Kang-PRL16, WangQH-PRB16, Chubukov-PRL12, Johnston-NJP16, Kontani-PRL10, Kontani-PRL12, Kontani-PRB17, Hirschfeld-PRB15, Hirschfeld-PRL16}.
However, resonant inelastic x-ray scattering measurements reveal that spin excitations in monolayer FeSe are gapped and nearly dispersionless, in contrast to the low-energy dispersive spin fluctuations characteristic of bulk FeSe \cite{Comin-NC21}. Although finite-energy spin fluctuations may still contribute to pairing \cite{WangJ-NL19, Weinert-CP18, WangJ-NL22}, these observations motivate the search for alternative electronic pairing mechanisms. 

Unlike the cuprates, whose parent compounds are charge-transfer insulators, FeSCs are $p$-$d$ systems with strong hybridization between transition-metal $d$ and ligand $p$ orbitals \cite{Imada-JPSJ10}, producing charge-transfer metals in which low-energy charge fluctuations can play an important role \cite{ZS-PRB11}. 
The intersite Coulomb repulsion between Fe $3d$ and Se $4p$ electrons can enhance collective charge-transfer fluctuations, characterized by out-of-phase charge fluctuations between the Fe and Se orbitals \cite{ZS-PRB11, Ono-JPSJ14, Ono-JPSJ18}. 
Such fluctuations are naturally enhanced at small momentum and can generate attractive pairing interactions in the spin-singlet channel. 
Moreover, the forward-scattering optical phonons of the SrTiO$_3$ substrate couple to charge degrees of freedom and may provide a natural complementary pairing interaction \cite{ShenZX-Nat14, ShenZX-PRL17, Johnston-NJP16}. 
In the present work, however, we focus on the consequences of enhanced electronic charge fluctuations themselves due to the intersite Coulomb interaction ($V_{pd}$) between the $d$ electrons of Fe and $p$ electrons of Se/Te.

In this work, we investigate charge-transfer-fluctuation-mediated superconductivity in 1L-FeSe/STO using an extended multiorbital Hubbard model incorporating all Fe $3d$ and Se $4p$ orbitals, onsite interactions on Fe, and intersite Fe-Se Coulomb repulsion. 
Since superconductivity is a Fermi-surface instability, we employ the random-phase approximation (RPA), a well-established weak-coupling approach, to examine the electronic fluctuation-mediated pairing interactions and identify the leading SC instability. 
We show that the intersite interaction strongly enhances long-wavelength $p$-$d$ charge-transfer fluctuations, which mediate attractive intraorbital spin-singlet pairing interactions predominantly within each electron pocket. 
Together with residual repulsive interpocket interactions, these fluctuations stabilize a nodeless $B_{1g}$ ($d$-wave) SC state with opposite gap signs on the two electron pockets \cite{Agterberg-PRL17}. Both Fe $t_{2g}$ and Se $4p$ orbitals participate substantially in the pairing. 
We studied both the mirror-glide-symmetric case and the case in which mirror-glide symmetry is broken by the substrate with the two-Fe unit cell, finding that the leading pairing instability and its essential characteristics remain robust in both cases.
The resulting gap anisotropy, two-peak tunneling density of states, and half-lattice-period modulation of the Josephson tunneling conductance are consistent with key experimental observations \cite{XueQK-CPL12, ShenZX-PRL16, FengDL-NP15, SJTM}, providing a microscopic electronic mechanism for the recently observed nodeless $d$-wave superconductivity.

\begin{figure}
	\begin{center}
		\fig{3.4in}{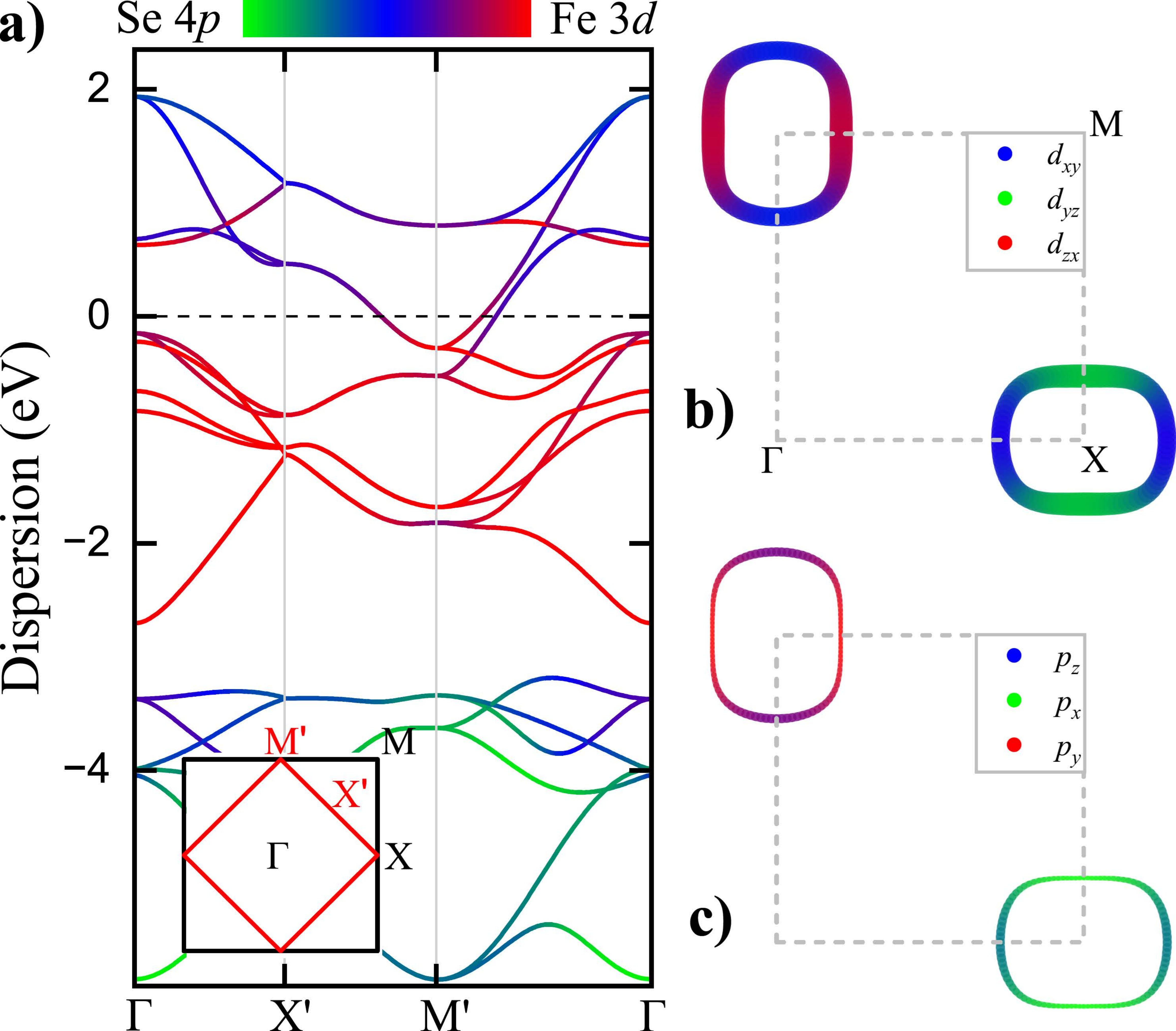}
		\caption{The $p$-$d$ tight-binding model for the FeSe layer in 1L-FeSe/STO. (a) Band dispersion in the reduced BZ, with the line color denoting Fe 3$d$ and Se 4$p$ contents. Inset shows the reduced two-Fe BZ (red lines) and the unfolded one-Fe BZ (black lines), with corresponding high-symmetry points labeled. (b) Fe $t_{2g}$ and (c) Se 4$p$ contributions to the FS in the unfolded BZ. The line thickness denotes the summed contribution from the three Fe $t_{2g}$ and Se 4$p$ orbitals, while the line color represents the content of each individual orbital.} \label{fig1}
	\end{center}
	\vskip-0.8cm
\end{figure}

\textit{Model and method.} -
We consider the extended Hubbard model as the low-energy effective description for the FeSe layer in 1L-FeSe/STO. The Hamiltonian is given by $\hat{H} = \hat{H}_0 +\hat{U}_{dd} +\hat{V}_{pd}$.
Here, $\hat{H}_0$ is a realistic two-dimensional tight-binding model including both Fe 3$d$ and Se 4$p$ orbitals,
\begin{align}
	\hat{H}_0 = &\sum_{i\alpha\sigma} \epsilon^d_\alpha d^\dagger_{i\alpha\sigma} d_{i\alpha\sigma} +\sum_{k\mu\sigma} \epsilon^p_\mu p^\dagger_{k\mu\sigma} p_{k\mu\sigma}  \label{H0} \\
	+&\sum_{ij,\alpha\beta,\sigma} t^{dd}_{i\alpha,j\beta} d^\dagger_{i\alpha\sigma} d_{j\beta\sigma} +\sum_{kl,\mu\nu,\sigma} t^{pp}_{k\mu,l\nu} p^\dagger_{k\mu\sigma} p_{l\nu\sigma} \nonumber \\
	+&\sum_{ik,\alpha\mu,\sigma} t^{pd}_{i\alpha,k\mu} d^\dagger_{i\alpha\sigma} p_{k\mu\sigma} +\text{h.c.} + \hat{H}_\text{sub}. \nonumber
\end{align}
Here $d^\dagger_{i\alpha\sigma}$ creates a spin-$\sigma$ electron on Fe site $i$ in orbital $\alpha$ ($\alpha,\beta$ = 1, 2, $\cdots$, 5 for $d_{xy}$, $d_{yz}$, $d_{zx}$, $d_{x^2-y^2}$, and $d_{3z^2-r^2}$) and $p^\dagger_{k\mu\sigma}$ creates a spin-$\sigma$ electron on Se site $k$ in orbital $\mu$ ($\mu,\nu$ = 6, 7, 8 for $p_z$, $p_x$, and $p_y$).
The onsite energies $\epsilon^d_\alpha$, $\epsilon^p_\mu$ and hopping parameters $t^{dd}_{i\alpha,j\beta}$, $t^{pp}_{k\mu,l\nu}$, and $t^{pd}_{i\alpha,k\mu}$ are given explicitly in the Supplemental Material (SM) \cite{supp}.
Importantly, $\hat{H}_\text{sub}$ in Eq. (\ref{H0}) represents the modification in electronic structure of the FeSe layer due to the presence of SrTiO$_3$ substrate \cite{Zhang-SR13, FengDL-NC14, XueQK-2DM16, SiQM-PRB15}.
Depending on the position of the substrate atoms with respect to the monolayer FeSe lattice, experiments have revealed two types of interfacial structure \cite{Ahn-SA20, CenC-arXiv25}.

\begin{figure*}
	\begin{center}
		\fig{7.in}{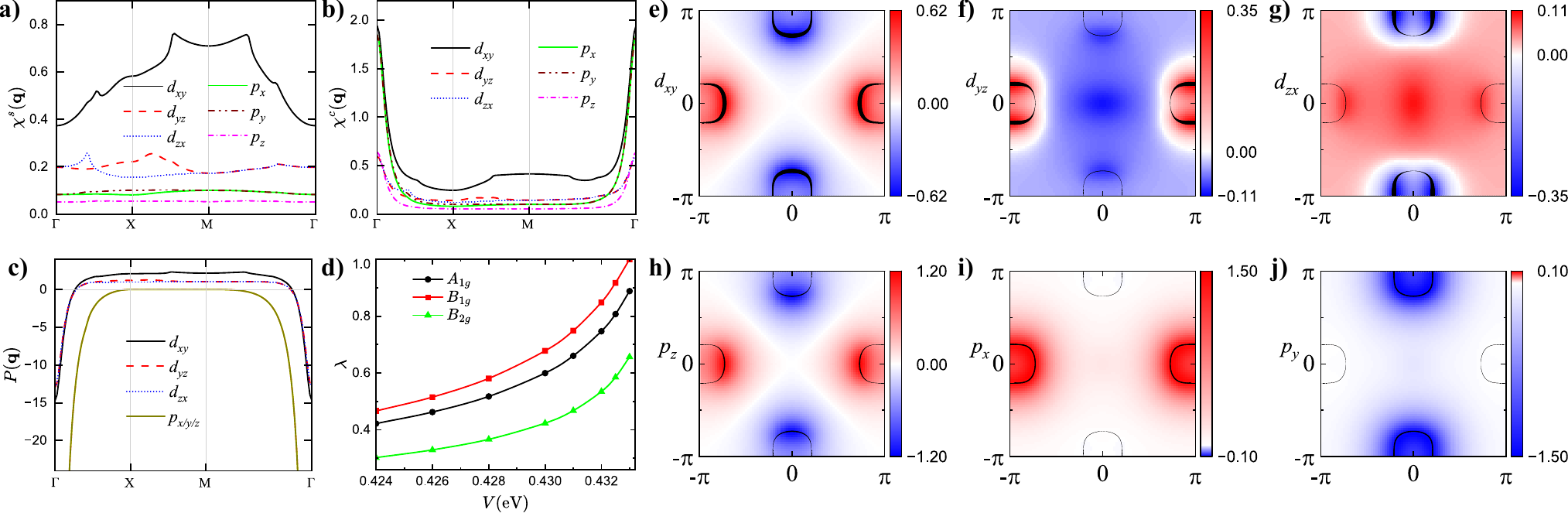}
		\caption{Intraorbital RPA (a) spin and (b) charge susceptibilities and (c) singlet pairing interactions obtained at $V$ = 0.433 eV. (d) The eigenvalues $\lambda$ of the linearized gap equation as a function of $V$. (e-j) The dominant intraorbital components of the normalized $B_{1g}$ gap function in the unfolded BZ.
			The on-site interactions are $J/U$ = 0.1 and $U$ = 1.2 eV.} \label{fig2}
	\end{center}
	\vskip-0.8cm
\end{figure*}

In the first type, the inversion and glide symmetries of the FeSe monolayer are broken, whereas the fourfold rotation $C_{4z}$ about a Se site that exchanges the two Fe sublattices is preserved.
In the second interface, the fourfold symmetry is also broken, rendering the two Fe sublattices inequivalent.
Since there is no clear signature of inequivalent Fe sites in Ref. \cite{SJTM}, we consider only the first type of interface in this work, while leaving the second to further study.
A minimal representation of this inversion asymmetry is a staggered onsite energy on Se atoms. We therefore write $\hat{H}_\text{sub}$ = $\varepsilon_p \sum_{k\mu\sigma} \eta_k p^\dagger_{k\mu\sigma} p_{k\mu\sigma}$ for the first type of interface, with $\eta_k$ = $\pm1$ for the lower and upper Se atoms.
The evolution of electronic structure and corresponding FS is presented in the SM as a function of the Se-site energy imbalance $\varepsilon_p$.
Within the explored parameter range, the FS consists of electron pockets only and, more importantly, the electron-pocket crossings along the edge of the reduced BZ remain ungapped at finite $\varepsilon_p$ \cite{supp}.
This is consistent with ARPES measurements that resolve the crossing pockets without a hybridization gap within the experimental resolution \cite{ShenZX-PRL16, Vafek-PRB18}.

Since the electron-pocket-only FS topology and the pocket crossings persist at nonzero $\varepsilon_p$, this suggest that the results with and without $\varepsilon_p$ may be qualitatively similar.
Therefore, we first consider the simplified case with $\varepsilon_p$ = 0, and then discuss the effects of the staggered onsite energy on Se atoms.
In the absence of the latter, the tight-binding Hamiltonian $\hat{H}_0$ recovers the glide symmetry and therefore resembles the free-standing monolayer FeSe.
Importantly, the recovery of the glide symmetry enables us to unfold the two-Fe unit cell into one-Fe unit cell in the absence of SOC \cite{supp, SiQM-PRB15}, where the physics is much more transparent.
The band structure of the $p$-$d$ model with $\varepsilon_p = 0$ is displayed in Fig. \ref{fig1}(a), where the orbital-resolved dispersion captures the principal features of the local-density-approximation (LDA) results \cite{Imada-JPSJ10}.
To account for electron transfer from the substrate, we electron-dope the model by 20\% \cite{supp}. The resulting unfolded FS consists only of electron pockets at $(\pi,0)$ and $(0,\pi)$ shown in Figs.~\ref{fig1}(b) and \ref{fig1}(c), with the $\Gamma$-centered hole bands below the Fermi level, consistent with ARPES \cite{ShenZX-Nat14,ShenZX-PRL16}.

The on-site interaction $\hat{U}_{dd}$ on Fe atoms is given by the standard multiorbital Hubbard model,
\begin{align}
	\hat{U}_{dd} = & U\sum_{i,\alpha} \hat{n}^d_{i\alpha\uparrow} \hat{n}^d_{i\alpha\downarrow} + \left(U' -\tfrac{1}{2} J \right) \sum_{i,\alpha<\beta} \hat{n}^d_{i\alpha} \hat{n}^d_{i\beta} \nonumber \\
	-& J\sum_{i,\alpha\neq\beta} {\bf S}^d_{i\alpha} \cdot {\bf S}^d_{i\beta} +J\sum_{i,\alpha\neq\beta} d^\dagger_{i\alpha\uparrow} d^\dagger_{i\alpha\downarrow} d_{i\beta\downarrow} d_{i\beta\uparrow}, \label{Udd}
\end{align}
with $U'$ = $U-2J$.
The density operators $\hat{n}^d_{i\alpha\sigma}$ = $d^\dagger_{i\alpha\sigma} d_{i\alpha\sigma}$, $\hat{n}^d_{i\alpha}$ = $\sum_\sigma \hat{n}^d_{i\alpha\sigma}$, and the spin operator ${\bf S}^d_{i\alpha}$ = $\tfrac{1}{2} \sum_{\sigma\sigma'} d^\dagger_{i\alpha\sigma} {\bm \tau}_{\sigma\sigma'} d_{i\alpha\sigma'}$ with ${\bm \tau} =(\tau_x, \tau_y, \tau_z)$ the vector of Pauli matrices.
The inter-site Coulomb interaction between neighboring Fe and Se atoms is given by
\begin{equation}
	\hat{V}_{pd} = V \sum_{\langle i,k\rangle,\alpha,\mu} \hat{n}^d_{i\alpha} \hat{n}^p_{k\mu}. \label{Vpd}
\end{equation}
Unlike the onsite interactions, $V$ couples Fe 3$d$ and Se 4$p$ charge density directly and promotes their relative charge fluctuations.
In the present work, it serves as an effective control parameter for charge fluctuations, irrespective of the microscopic mechanism responsible for their enhancement.

Next, we calculate the RPA spin and charge susceptibilities, and investigate the leading pairing instability by solving the corresponding linearized gap equation.
The RPA spin and charge susceptibilities in the presence of interactions
$\hat{U}_{dd}$ and $\hat{V}_{pd}$ can be written as $28\times 28$ matrices in the interaction-coupled orbital-pair basis \cite{supp}
\begin{align}
	\hat{\chiup}^s (\bq, \omega_\ell) &= \hat{\chiup}^0 (\bq, \omega_\ell) \{1- \hat{U}^s \hat{\chiup}^0 (\bq, \omega_\ell) \}^{-1}, \\
	\hat{\chiup}^c (\bq, \omega_\ell) &= \hat{\chiup}^0 (\bq, \omega_\ell) \{ 1+ [\hat{U}^c +2\hat{V}^c (\bq)] \hat{\chiup}^0 (\bq, \omega_\ell) \}^{-1},
\end{align}
where the bare susceptibilities $\chiup^0_{ab,a'b'} (\bq, \omega_\ell)$ = $-(T/N) \sum_{\bk, m}$ $G^0_{aa'} (\bk+\bq, \epsilon_m +\omega_\ell)$ $G^0_{b'b} (\bk, \epsilon_m)$ with the noninteracting Green's function $\hat{G}^0 (\bk, \epsilon_m) =[i\epsilon_m -\hat{H}_0 (\bk)]^{-1}$.
The nonzero elements of the interaction matrices $\hat{U}^s$, $\hat{U}^c$, and $\hat{V}^c$ are $U^s_{\alpha\alpha, \alpha\alpha}$ = $U$, $U^s_{\alpha\beta, \alpha\beta}$ = $U'$, $U^s_{\alpha\alpha, \beta\beta}$ = $J$, $U^s_{\alpha\beta, \beta\alpha}$ = $J$, $U^c_{\alpha\alpha, \alpha\alpha}$ = $U$, $U^c_{\alpha\beta, \alpha\beta}$ = $2J -U'$, $U^c_{\alpha\alpha, \beta\beta}$ = $2U' -J$, $U^c_{\alpha\beta, \beta\alpha}$ = $J$, and $V^c_{\alpha\alpha, \mu\mu} (\bq)$ = $V(\bq)\equiv 4V \cos{\frac{1}{2}q_x} \cos{\frac{1}{2}q_y}$.
Clearly, the on-site interactions enhance the spin susceptibility while suppressing the charge susceptibility.
In contrast, the intersite interaction $V$ contributes only to the charge channel, entering $\hat{\chi}^c$ through off-diagonal Fe-Se blocks in the denominator.

To study superconductivity, we evaluate the pairing vertex dressed by the spin and charge fluctuations.
The effective spin-singlet pairing interaction is given by
\begin{align}
	\hat{P}(\bq)=& \tfrac{1}{2} [\hat{U}^s +\hat{U}^c +2\hat{V}^c(\bq)] + \tfrac{3}{2} \hat{U}^s \hat{\chiup}^s(\bq) \hat{U}^s  \nonumber \\
	-&\tfrac{1}{2} [\hat{U}^c +2\hat{V}^c(\bq)] \hat{\chiup}^c (\bq) [\hat{U}^c +2\hat{V}^c(\bq)], \label{pairing}
\end{align}
where $\hat{\chiup}^{s,c} (\bq)=\hat{\chiup}^{s,c} (\bq,\omega_l=0)$ are the static spin and charge susceptibilities.
The spin-triplet pairing turns out to be subleading.
The superconducting instability is determined by solving the linearized gap equation,
\begin{align}
	\lambda \Delta_{ab} (\bk) = -\frac{T}{N} & \sum_{\bk',n} \sum_{a'b',a''b''} P_{aa'',b''b} (\bk-\bk') \nonumber \\
	\times & G^0_{a''a'}(\bk',\omega_n) G^0_{b''b'}(-\bk',-\omega_n) \Delta_{a'b'} (\bk'), \label{gapequation}
\end{align}
where $\Delta_{ab}(\bk)$ denotes the $8\times 8$ normalized gap function in the orbital basis of the unfolded one-Fe unit cell.
Within the linearized treatment, the superconducting instability occurs when the largest eigenvalue $\lambda$ reaches unity.
To minimize finite-size effects, Eq.~(\ref{gapequation}) is solved self-consistently on an $80\times80$ momentum mesh at $T$= 20 meV for various interaction parameters.
The gap structure on FS is then obtained by transforming the orbital-basis gap function $\Delta_{ab}(\bk)$ into the band basis $\Delta_n(\bk)$.

\textit{Nodeless $d$-wave superconductivity driven by charge fluctuations.} -
We fix the on-site interactions on Fe sites to $U$ = 1.2 eV and $J/U$ = 0.1, and investigate superconductivity driven primarily by charge fluctuations enhanced by the intersite Coulomb repulsion $V$ between neighboring Fe and Se atoms.
The intraorbital static spin and charge susceptibilities $\chiup^{s, c}_{aa, aa} (\bq)$ are shown in Figs. \ref{fig2}(a) and \ref{fig2}(b), respectively.
The spin susceptibilities are independent of $V$ and exhibit a broad peak around ($\pi$, $\pi$) in the intraorbital $d_{xy}$ component.
In contrast, the intraorbital charge fluctuations are strongly enhanced by $V$, generating a pronounced peak at $\bq$ = (0, 0), where $V(\bq)$ reaches its maximum, as shown in Fig. \ref{fig2}(b).
Consequently, the repulsive intraorbital pairing potential is substantially reduced and eventually becomes attractive near $\bq$ = (0, 0) when $V$ is sufficiently strong, as displayed in Fig. \ref{fig2}(c).
This behavior is qualitatively different from intraorbital contributions of spin fluctuations, which remain repulsive at all momenta.
Interestingly, the intraorbital pairing potentials of the three Se 4$p$ orbitals
$\hat{P}_{\mu\mu,\mu\mu}(\bq)$ = $-2\sum_{\alpha,\beta} \hat{V}^c_{\mu\mu, \alpha\alpha}(\bq) \hat{\chi}^c_{\alpha\alpha, \beta\beta} (\bq) \hat{V}^c_{\beta\beta, \mu\mu}(\bq)$
involve exclusively the charge fluctuations of Fe 3$d$ orbitals, mediated by the intersite interactions.
They are identical and become the most attractive components at sufficiently large $V$.

\begin{figure}
	\begin{center}
		\fig{3.4in}{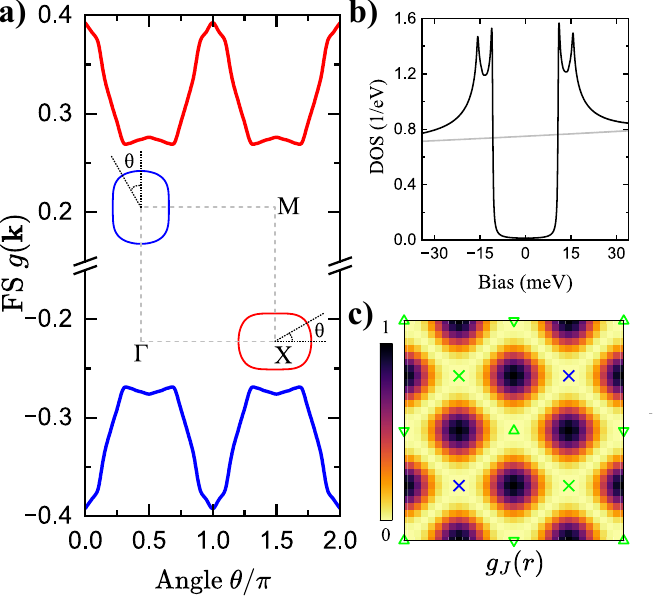}
		\caption{(a) The normalized gap function on the electron pockets at $V_c$, where the $B_{1g}$ eigenvalue $\lambda$ reaches 1. The inset defines the angles. (b) DOS of the $B_{1g}$ superconductivity with pairing strength $\Delta_0$ = 0.1 eV, in comparison with the normal-state DOS (gray line). (c) Spatial distribution of the normalized Josephson tunneling conductance $g_J(\br)$, superimposed with the positions of Fe$_A$ (green cross), Fe$_B$ (blue cross), Se$_A$ (up triangle), and Se$_B$ (down triangles) atoms.} \label{fig3}
	\end{center}
	\vskip-0.8cm
\end{figure}

The pairing symmetry obtained via solving Eq. (\ref{gapequation}) has contributions from all orbitals and can be qualitatively understood from the intraorbital pairing interactions displayed in Fig. \ref{fig2}(c).
The attractive interactions peaked at small momentum provide the primary pairing force via forward scattering, which strongly disfavors any sign change in the superconducting gap on each individual FS pocket.
At the same time, the residual repulsive interactions around ($\pi$, $\pi$), originating mainly from spin fluctuations, favor opposite signs of the gap functions on the two electron pockets centered at ($\pi$, 0) and (0, $\pi$).
These two tendencies cooperatively stabilize a nodeless $B_{1g}$ superconducting state with opposite gap signs on the two electron pockets.
Indeed, the eigenvalues $\lambda$ plotted as a function of $V$ in Fig. \ref{fig2}(d) show that the leading pairing instability is toward a nodeless $B_{1g}$ ($d$-wave) superconductivity, while the nodeless $A_{1g}$ ($s$-wave) solution with the same gap sign on the two pockets is subleading, and the $B_{2g}$ ($d$-wave) state with sign changes on both electron pockets is least favorable.
For comparison, we also examine the $V=0$ limit, where spin fluctuations yield a leading nodeless $B_{1g}$ instability at small $J/U$ and an $A_{1g}$ instability at larger $J/U$ (see SM \cite{supp}).
At larger $J/U$, increasing $V$ enhances small-momentum charge fluctuations and changes the leading pairing instability from $A_{1g}$ to $B_{1g}$.

The $B_{1g}$ superconductivity sets in at a moderate critical value $V_c$ = 0.433 eV, with the dominant intraorbital components of the normalized gap functions $\Delta_{aa} (\bk)$ presented in Figs. \ref{fig2}(e-j) for the three Fe $t_{2g}$ orbitals and the three Se 4$p$ orbitals, together with the corresponding orbital weights on the FS.
Remarkably, the gap functions are concentrated on the electron pockets and closely track the associated orbital character.
In particular, the weights of the Fe $t_{2g}$ and Se $p_z$ orbitals vary substantially on the pockets, and the corresponding intraorbital gap functions exhibit pronounced anisotropy.
Explicitly, the maxima of the $d_{xy}$ and $p_z$ gap functions are located near the vertices of the elliptical electron pockets, while those of the $d_{yz/zx}$ orbitals occur near the co-vertices, as shown in Fig. \ref{fig2}(e-h).
The combined effects of the orbital-dependent gap functions and orbital contents give rise to the FS gap structure displayed in Fig. \ref{fig3}(a), whose pronounced anisotropy -- with two maxima near the major axes of the pockets and minima near their intersections -- is consistent with ARPES measurements \cite{ShenZX-PRL16}.

To make further connections to STM and SJTM measurements, we introduce the phenomenological Hamiltonian
\begin{equation}
	\hat{H}_p=\hat{H}_0 + \Delta_0 \sum_{\bk, ab} \Delta_{ab}(\bk) \left(c_{a\bk\uparrow} c_{b\bar{\bk} \downarrow} -c_{a\bk \downarrow} c_{b\bar{\bk}\uparrow} \right) +\text{H.c.},
\end{equation}
where $\Delta_{ab}(\bk)$ is the normalized $B_{1g}$ gap function obtained above, and $\Delta_0$ controls the overall pairing strength.
The resulting superconducting density of states (DOS) for $\Delta_0$ = 0.1 eV is presented in Fig. \ref{fig3}(b).
It exhibits a U-shaped double-gap structure qualitatively consistent with the tunneling spectra observed in STM \cite{XueQK-CPL12, FengDL-NP15}.
We then calculate the spatially resolved Josephson tunneling conductance $g_J(\br)$ between an $s$-wave superconducting tip and the nodeless $B_{1g}$ superconductivity described by $\hat{H}_p$, retaining only the lowest-harmonic contribution, see SM \cite{supp} for details.
As shown in Fig. \ref{fig3}(c), $g_J(\br)$ exhibits a half-lattice-period modulation, reaching its maximum at the midpoints of NN Fe-Fe bonds while being completely suppressed along next-NN Fe-Fe directions, resembling closely the pattern observed in SJTM \cite{SJTM}.
We emphasize that neither the FS gap amplitude nor the superconducting DOS can unambiguously distinguish between nodeless $A_{1g}$ and nodeless $B_{1g}$ pairing \cite{supp, Vafek-PRB18}.
In contrast, the Josephson tunneling conductance is sensitive to the gap's relative phase and therefore provides phase-sensitive support for the $B_{1g}$ ($d$-wave) interpretation in 1L-FeSe/STO.

\begin{figure}
\begin{center}
\fig{3.4in}{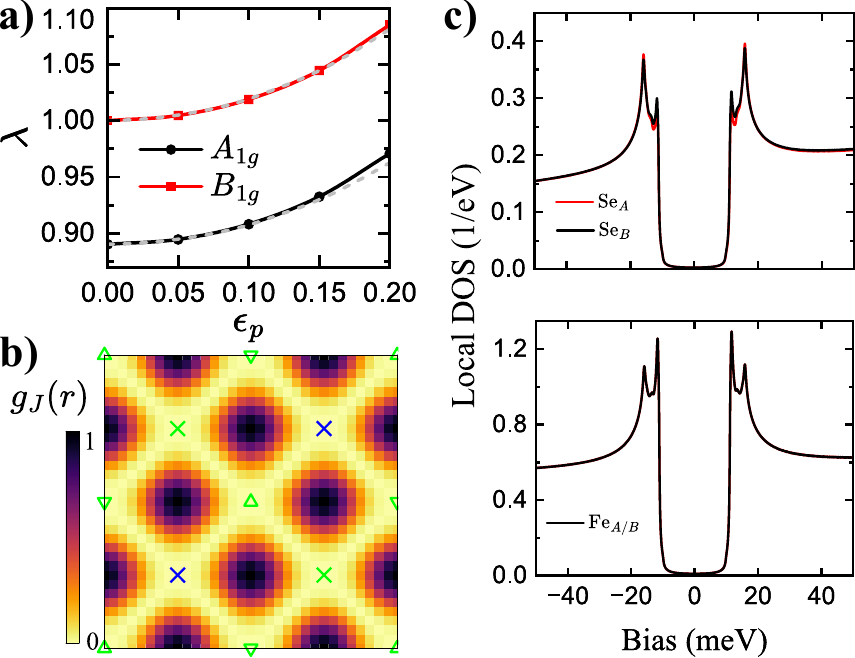}
\caption{(a) The pairing eigenvalues $\lambda$ of the $A_{1g}$-derived and $B_{1g}$-derived channels, continuously connected to the $A_{1g}$ and $B_{1g}$ representations at $\varepsilon_p$ = 0, as functions of the substrate-induced Se-site energy imbalance $\varepsilon_p$.
The gray curves show $\lambda(\varepsilon_p)$ = $\lambda(0) \exp [a/N_F(0) -a/N_F(\varepsilon_p)]$, with constant $a$ = 170 eV$^{-1}$ and $N_F(\varepsilon_p)$ the normal-state Fermi-level DOS at $\varepsilon_p$. (b) Spatial distribution of the normalized Josephson tunneling conductance $g_J(\br)$, superimposed with the positions of Fe and Se atoms. (c) The local DOS on the two inequivalent Se sublattices and the two symmetry-equivalent Fe sites.} \label{fig4}
\end{center}
\vskip-0.8cm
\end{figure}

\textit{Effects of staggered onsite energy on Se atoms.} -
Next, we perform the calculations in the two-Fe unit cell. 
The substrate-induced Se-site energy imbalance $\varepsilon_p$ breaks inversion and glide symmetries while preserving the $C_{4v}$ point-group symmetry. 
Consequently, the $A_{1g}$ and $B_{1g}$ pairing representations of the centrosymmetric system reduce to $A_1$ and $B_1$, respectively. 
In the following, we refer to these states as the $A_{1g}$-derived and $B_{1g}$-derived pairing channels to emphasize their continuity with the corresponding states at $\varepsilon_p=0$. 
In the absence of $\varepsilon_p$, the consistency between calculations in two- and one-Fe unit cells is verified numerically, as illustrated in the SM \cite{supp}. 
We then switch on the staggered onsite energy on Se atoms at fixed $U=1.2$ eV, $J/U=0.1$, and $V=0.433$ eV, with the eigenvalues $\lambda$ displayed in Fig.~\ref{fig4}(a) as a function of $\varepsilon_p$. 
Both eigenvalues grow monotonically over the explored range, closely tracking the increase in the normal-state Fermi-level DOS. 
Importantly, the leading pairing instability remains in the $B_{1g}$-derived channel throughout the explored range of $\varepsilon_p$, demonstrating the robustness of the leading nodeless $d$-wave pairing instability against substrate-induced inversion and glide symmetry breaking.

Fig. \ref{fig4}(b) displays the spatially resolved Josephson tunneling conductance $g_J(\br)$ obtained in the two-Fe unit-cell calculation at $\varepsilon_p$ =0.2 eV.
Its overall structure agrees with that obtained in the absence of $\varepsilon_p$ [Fig.~\ref{fig3}(c)], since the zero-lines along next-NN Fe-Fe directions are strictly enforced by the $B_1$ symmetry of the superconducting phase.
Nevertheless, a nonzero $\varepsilon_p$ breaks the inversion and glide symmetry, thus the two Se sublattices are no longer symmetry related and become inequivalent.
This is visualized by the small difference between their local DOS displayed in Fig. \ref{fig4}(c) for $\varepsilon_p$ = 0.2 eV and $\Delta_0$ = 0.1 eV.
On the other hand, the fourfold rotation about a Se site that exchanges the two Fe sublattices is preserved.
The local DOS on the two Fe sites therefore remains identical, as shown in Fig.~\ref{fig4}(c).

\textit{Summary and discussion.} -
We have investigated superconductivity in 1L-FeSe/STO within an extended multiorbital Hubbard model incorporating both Fe 3$d$ and Se 4$p$ orbitals.
Motivated by the suppression of low-energy dispersive spin fluctuations \cite{Comin-NC21} and the growing evidence for the importance of charge and orbital degrees of freedom in FeSe-derived superconductors \cite{Zhao-SA18, PengYY-npjQM22}, we explored an electronic pairing mechanism meidated by Fe-Se charge-transfer fluctuations.
Within the RPA and the linearized gap equation, we find that the intersite Fe-Se Coulomb repulsion strongly enhances long-wavelength $p$-$d$ charge fluctuations, generating attractive spin-singlet pairing interactions predominantly within each electron pocket.
Together with residual repulsive interpocket interactions, these fluctuations favor a robust nodeless $B_{1g}$ ($d$-wave) SC state with opposite gap signs on the two electron pockets.

We further demonstrate that the leading $d$-wave instability remains robust against the substrate-induced Se-site energy imbalance $\varepsilon_p$, which breaks inversion and glide symmetries while preserving the $C_{4v}$ point-group symmetry. The resulting $B_{1g}$-derived pairing state is continuously connected to the $B_{1g}$ state of the centrosymmetric system and remains nodeless on the electron-pocket Fermi surfaces within our SOC-free model \cite{ShenZX-Nat14, ShenZX-PRL16}. The calculated gap anisotropy, two-peak tunneling density of states, and half-lattice-period modulation of the Josephson tunneling conductance are consistent with key experimental observations, including recent scanning Josephson tunneling microscopy measurements \cite{ShenZX-PRL16, XueQK-CPL12, FengDL-NP15, SJTM}. Our results thus identify Fe-Se charge-transfer fluctuations as a viable microscopic mechanism for nodeless $d$-wave superconductivity in 1L-FeSe/STO.

The leading SC instability eigenvalue reaches $\lambda$ = 1 at an intersite interaction $V_c\approx0.43$ eV for the chosen parameters at $T=20$ meV, indicating that a SC instability can develop at a moderate effective interaction strength within our RPA treatment. 
This scale is comparable to the Fe-pnictogen intersite Coulomb interaction considered in earlier charge-fluctuation theories of FeSC \cite{ZS-PRB11}, although it should not be interpreted as a material-specific estimate for 1L-FeSe/STO.
First-principles studies of bulk FeSe also suggest that nonlocal Fe-Se Coulomb interactions can remain substantial after electronic screening \cite{Imada-JPSJ10}, while other theoretical studies have highlighted the importance of nonlocal interactions for its electronic structure \cite{Hirschfeld-PRB20, Jiang-PRB16}.
An additional feature is the coupling to optical phonons of the SrTiO$_3$ substrate, whose predominantly forward-scattering character may provide a complementary attractive pairing interaction \cite{GuoJD-PRB16, FengDL-NC19, ShenKM-PRL21, PanXQ-Nat24}.
Since both the charge-transfer fluctuations considered here and the interfacial phonons favor small-momentum scattering, their cooperative effects may further enhance intrapocket pairing without necessarily altering the relative sign structure of the SC order parameter between the electron pockets.
In the present work, we treat the intersite Coulomb interaction as an effective parameter controlling the strength of charge-transfer fluctuations and focused on the resulting leading SC instability. A quantitative determination of the screened Fe-Se interaction, together with a microscopic treatment of its interplay with interfacial phonons, remains an important direction for future work \cite{Oppeneer-PRB20, FengDL-NL24}.

Finally, we discuss the implications of our results for the second type of substrate interface that additionally breaks the fourfold rotation symmetry relating the two Fe sublattices.
Such symmetry breaking can be modeled by introducing a staggered Fe-site potential $\varepsilon_d$, in addition to the Se-site imbalance $\varepsilon_p$.
In the crystallographic convention of Refs. \cite{Agterberg-arXiv25, HuJP-arXiv25},
these perturbations transform as $B_{2u}$ and $A_{2u}$, respectively.
Unlike $\varepsilon_p$, which preserves the equivalence of the two Fe sublattices, $\varepsilon_d$ removes the symmetry protecting their identical local electronic spectra and can therefore allow the particle--hole-antisymmetric Fe-sublattice contrast reported by STM \cite{XueQK-PRL26}.
In the nodeless $d$-wave scenario, the resulting reconstruction of the electronic bands can also generate interband pairing components in the perturbed band basis, providing a connection between symmetry-based and interband-pairing descriptions of the sublattice dichotomy \cite{Agterberg-arXiv25, HuJP-arXiv25}.
By contrast, the Se-site imbalance considered explicitly in this work produces only a small difference between the local Se-site spectra while leaving the Fe-site spectra equivalent [Fig.~\ref{fig4}(c)].
Our calculations further show that the electron-pocket crossings along the edge of the reduced BZ remain ungapped under the Se-site imbalance, whereas a staggered Fe-site potential opens a gap at these crossings. SOC provides another source of electron-pocket hybridization \cite{Vafek-PRB13, Vafek-PRB18, Agterberg-PRL17, Agterberg-PRB18}.
The absence of a clearly resolved avoided crossing in ARPES measurements places constrains on the effective coupling between the electron-pocket states \cite{ShenZX-PRL16,Vafek-PRB18}.
Previous multiband studied have shown that nodeless $d$-wave superconductivity can survive sufficiently weak SOC-induced hybridization when interband pairing is properly included  \cite{Agterberg-PRL17}.
Extending the present charge-transfer-fluctuation framework to incorporate both staggered Fe-site potentials and SOC would therefore be valuable for establishing the general conditions for robust nodeless $d$-wave superconductivity and the Fe-sublattice dichotomy in 1L-FeSe/STO.

\textit{Acknowledgments.} -
We thank Prof. Xianxin Wu for stimulating discussions.
This work is supported by the National Key Research and Development Program of China (Grant No. 2022YFA1403800), the National Natural Science Foundation of China (Grant Nos. 12374153 and 12447101), the Innovation Program for Quantum Science and Technology (Grant No. 2024ZD0300500), and the Postdoctoral Fellowship Program of CPSF (Grant No. GZC20241749).
Z.W. is supported by the U.S. Department of Energy, Basic Energy Sciences
(Grant No. DE-FG02-99ER45747).
Numerical calculations in this work were performed on the HPC Cluster of ITP-CAS.

\bibliography{ref}
\clearpage
\end{document}